\documentclass[
reprint,
superscriptaddress,
showpacs,
amsmath,amssymb,
aps,
prc,
floatfix,
altaffilletter,
]{revtex4-1}
\usepackage[T1]{fontenc}
\usepackage[utf8]{inputenc}
\usepackage{todonotes}
\usepackage{graphicx}
\usepackage{dcolumn}
\usepackage{bm}
\usepackage{hyperref}
\usepackage{float}
\usepackage{epstopdf}
\usepackage{color}
\usepackage{xcolor,colortbl}

\graphicspath{{fig/}}

\begin{document}

\preprint{APS/123-QED}

\title{The structure of low-lying states in \texorpdfstring{${}^{140}$Sm}{140Sm} studied by Coulomb excitation}

\newcommand{\afOslo}{Department of Physics, University of Oslo, N-0316 Oslo, Norway}
\newcommand{\afLodz}{Faculty of Physics, University of Lodz, 90-236 Lodz, Poland}
\newcommand{\afWarsawFac}{Faculty of Physics, University of Warsaw, 02-093 Warsaw, Poland}
\newcommand{\afWarsawLab}{Heavy Ion Laboratory, University of Warsaw, 02-093 Warsaw, Poland}
\newcommand{\afDarmstadt}{Institut f\"ur Kernphysik, Technische Universit\"at Darmstadt, Darmstadt D-64289, Germany}
\newcommand{\afGeneva}{University of Geneva, Bd du Pont-d'Arve 40, 1211 Genéve, Switzerland}
\newcommand{\afOsloChem}{Department of Chemistry, University of Oslo, N-0316 Oslo, Norway}
\newcommand{\afSaclay}{CEA Saclay, IRFU, SPHN, F-91191 Gif-sur-Yvette, France}
\newcommand{\afSwierk}{National Centre for Nuclear Research, \'Swierk, 05-400 Otwock, Poland}
\newcommand{\afMunchen}{Technische Universit\"at M\"unchen, Garching D-85748, Germany}
\newcommand{\afJyvaskyla}{University of Jyv\"askyl\"a, Department of Physics, P.O. Box 35, FI-40014 University of Jyv\"askyl\"a, Finland}
\newcommand{\afHelsinki}{Helsinki Institute of Physics, University of Helsinki, P.O. Box 64, FIN-00014 Helsinki, Finland}
\newcommand{\afLeuven}{Instituut voor Kern- en Stralingsfysica, KU Leuven, Leuven B-3001, Belgium}
\newcommand{\afIsolde}{CERN, CH-1211 Geneva 23, Switzerland}
\newcommand{\afKoln}{Institut f\"{u}r Kernphysik, Universit\"{a}t zu K\"{o}ln, K\"{o}ln D-50937, Germany}
\newcommand{\afYork}{Department of Physics, University of York, York YO10 5DD, UK}
\newcommand{\afCEAetc}{CEA, DAM, DIF, F-91297 Arpajon, France}
\newcommand{\afIPHCa}{IPHC, Universit\'e de Strasbourg, IPHC, 23 rue du Loess 67037 Strasbourg, France}
\newcommand{\afIPHCb}{IPHC, CNRS, UMR7178, 67037 Strasbourg, France}
\newcommand{\afHelmholtz}{Helmholtz Institute Mainz, 55099 Mainz, Germany}
\newcommand{\afGSI}{GSI Helmholtzzentrum f\"{u}r Schwerionenforschung, 64291 Darmstadt, Germany}
\newcommand{\afLNL}{INFN, Laboratori Nazionali di Legnaro, I-35020 Legnaro, Italy}
\newcommand{\afGutenberg}{Institut f\"{u}r Physik, Johannes Gutenberg-Universit\"{a}t Mainz, D-55128 Mainz, Germany}
\newcommand{\afPetersburg}{Petersburg Nuclear Physics Institute, NRC Kurchatov Institute, Gatchina 188300, Russia}
\newcommand{\afKEK}{High Energy Accelerator Research Organization (KEK), Oho 1-1, Tsukuba, Ibaraki 305-0801, Japan}

\author{M.~Klintefjord}\email{malin.klintefjord@fys.uio.no}
\affiliation{\afOslo}

\author{K.~Hady\'nska-Kl\c{e}k}\email{katarzyna.hadynska@lnl.infn.it}
\affiliation{\afOslo}
\affiliation{\afLNL}

\author{A.~G\"{o}rgen}\email{andreas.gorgen@fys.uio.no}
\affiliation{\afOslo}

\author{C.~Bauer}
\affiliation{\afDarmstadt}

\author{F.L.~Bello~Garrote}
\affiliation{\afOslo}

\author{S.~B\"onig}
\affiliation{\afDarmstadt}

\author{B.~Bounthong}
\affiliation{\afIPHCa}
\affiliation{\afIPHCb}

\author{A.~Damyanova}
\affiliation{\afGeneva}

\author{J.-P.~Delaroche}
\affiliation{\afCEAetc}

\author{V.~Fedosseev}
\affiliation{\afIsolde}

\author{D.A.~Fink}
\affiliation{\afIsolde}

\author{F.~Giacoppo}
\email{Presently at \afHelmholtz{}}
\affiliation{\afOslo}

\author{ M.~Girod}
\affiliation{\afCEAetc}

\author{P.~Hoff}
\affiliation{\afOsloChem}

\author{N.~Imai}
\affiliation{\afKEK}

\author{W.~Korten}
\affiliation{\afSaclay}

\author{A.C.~Larsen}
\affiliation{\afOslo}

\author{J.~Libert}
\affiliation{\afCEAetc}

\author{R.~Lutter}
\affiliation{\afMunchen}

\author{B.A.~Marsh}
\affiliation{\afIsolde}

\author{P.L.~Molkanov}
\affiliation{\afPetersburg}

\author{H.~Na\"{i}dja}
\affiliation{\afIPHCa}
\affiliation{\afIPHCb}

\author{P.~Napiorkowski}
\affiliation{\afWarsawLab}

\author{F.~Nowacki}
\affiliation{\afIPHCa}
\affiliation{\afIPHCb}

\author{J.~Pakarinen}
\affiliation{\afJyvaskyla}
\affiliation{\afHelsinki}

\author{E.~Rapisarda}
\affiliation{\afLeuven}
\affiliation{\afIsolde}

\author{P.~Reiter}
\affiliation{\afKoln}

\author{T.~Renstr\o{}m}
\affiliation{\afOslo}

\author{S.~Rothe}
\affiliation{\afIsolde}
\affiliation{\afGutenberg}

\author{M.D.~Seliverstov}
\affiliation{\afPetersburg}

\author{B.~Siebeck}
\affiliation{\afKoln}

\author{S.~Siem}
\affiliation{\afOslo}

\author{J.~Srebrny}
\affiliation{\afWarsawLab}

\author{T.~Stora}
\affiliation{\afIsolde}

\author{P.~Th\"ole}
\affiliation{\afKoln}

\author{T.G.~Tornyi}
\affiliation{\afOslo}

\author{G.M.~Tveten}
\affiliation{\afOslo}

\author{P.~Van~Duppen}
\affiliation{\afLeuven}

\author{M.J~Vermeulen}
\affiliation{\afYork}

\author{D.~Voulot}
\affiliation{\afIsolde}

\author{N.~Warr}
\affiliation{\afKoln}

\author{F.~Wenander}
\affiliation{\afIsolde}

\author{H.~De~Witte}
\affiliation{\afLeuven}

\author{M.~Zieli\'nska}
\affiliation{\afSaclay}

\date{\today}
\begin{abstract}
The electromagnetic structure of $^{140}$Sm was studied in a low-energy Coulomb excitation experiment with a radioactive ion beam from the REX-ISOLDE facility at CERN. The $2^+$ and $4^+$ states of the ground-state band and a second $2^+$ state were populated by multi-step excitation. The analysis of the differential Coulomb excitation cross sections yielded reduced transition probabilities between all observed states and the spectroscopic quadrupole moment for the $2_1^+$ state. The experimental results are compared to large-scale shell model calculations and beyond-mean-field calculations based on the Gogny D1S interaction with a five-dimensional collective Hamiltonian formalism. Simpler geometric and algebraic models are also employed to interpret the experimental data. The results indicate that $^{140}$Sm shows considerable $\gamma$ softness, but in contrast to earlier speculation no signs of shape coexistence at low excitation energy. This work sheds more light on the onset of deformation and collectivity in this mass region. 
\end{abstract}

\pacs{25.70.De,21.10.Ky,21.60.Ev,27.60.+j}
\maketitle

\section{Introduction}\label{sec:introduction}

The shape of atomic nuclei is a~fundamental property which is governed by the interplay between single-particle and collective degrees of freedom. Nuclei with closed proton and neutron shells are spherical in their ground state, whereas the occupation of shape-driving orbitals causes nuclei with open shells to be deformed. The majority of deformed nuclei is found to have prolate (elongated) quadrupole shapes. For heavy nuclei with $Z,N > 50$, oblate ground-state shapes are rare and mostly found in regions with holes in high-spin, low-$\Omega$ orbitals near the top of the proton and neutron shells~\cite{Bohr1975}. Mean-field calculations based on Gogny~\cite{Delaroche2010} or Skyrme~\cite{Dobaczewski2004} interactions predict oblate ground states, e.g.\ for platinum and mercury isotopes with $N > 106$ and for $N \approx 78$ and $N \approx 120$ isotones with $Z > 60$.

In certain regions of the nuclear chart, prolate and oblate shapes are found to coexist at low excitation energy. Shape coexistence at low energy can generally be expected when there is a~competition between an energy gap and a~residual interaction that favors excitations across the gap~\cite{Heyde2011,Wood2016}. Prominent examples for shape coexistence based on this mechanism are found in the $Z=82$ region near neutron mid-shell, e.g.\ in $^{186}$Pb~\cite{Andreyev2000} and the neighboring mercury isotopes~\cite{Bree2014}. Not surprisingly these nuclei lie in the region where mean-field calculations predict the transition from prolate to oblate ground-state shapes. From these observations one may expect to find shape coexistence also in the $N \approx 78$ region near proton mid-shell, where the role of protons and neutrons is interchanged. However, clear experimental indication for shape coexistence near the ground state is lacking in this region, and it was argued that the $Z=64$ sub-shell closure could explain the absence of shape coexistence in this region~\cite{Heyde2011}. There is, however, evidence for shape coexistence at higher excitation energy in the $N=78$ isotones $^{140}$Sm and $^{142}$Gd. Two isomeric $10^+$ states based on the configuration $({\pi}h_{11/2})^2$ and $({\nu}h_{11/2})^{-2}$ are found in both nuclei~\cite{Starzecki1988}. Lifetime measurements found the rotational bands built on top of the $10^+$ states with proton-particle and neutron-hole character in $^{140}$Sm to be consistent with prolate and oblate shape, respectively~\cite{Cardona1991}. The shape associated with the states in the ground-state band of $^{140}$Sm below the $10^+$ isomers is not clear and is the subject of the present investigation.

\begin{figure}[ht!]
	\centering
	\includegraphics[width=\columnwidth]{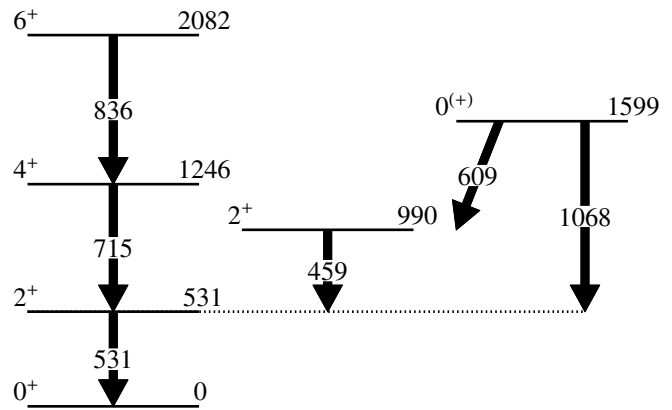}
	\caption{Partial level scheme showing low-lying states in ${}^{140}$Sm. The spin assignments for the excited states at 990~keV and 1599~keV are taken from a~recent angular correlation measurement~\cite{Samorajczyk2015}.}
	\label{fig:level}
\end{figure}

Relativistic mean-field calculations restricted to axial deformations predicted the ground state of $^{140}$Sm to have oblate shape, whereas prolate deformation was found to develop rapidly in the lighter samarium isotopes with $N \leq 76$~\cite{Lalazissis1996}. More recent relativistic Hartree Fock Bogoliubov (HFB) calculations find a~smooth transition from spherical $^{144}$Sm to well-deformed prolate $^{134}$Sm with a~$\gamma$-soft potential energy surface for the transitional nucleus $^{140}$Sm~\cite{Niksic2010}. The observation and tentative assignment of a~$(2_2^+)$ state at 990~keV and a~$(3_1^+)$ state at 1599~keV in $^{140}$Sm following the $\beta$ decay of $^{140}$Eu was interpreted as evidence for a~low-lying $\gamma$ band~\cite{Kern1987}. The observation was supported by triaxial-rotor calculations based on a~Woods-Saxon potential, which reproduced the excitation energies of the presumed $\gamma$ band and also found the potential energy surface for $^{140}$Sm to be soft in the triaxial degree of deformation~\cite{Kern1987}. However, a~subsequent $\beta$-decay experiment
revised the earlier spin values and tentatively assigned spin-parity $(0^+)$ and $(2^+)$ to the states at 990~keV and 1599~keV, respectively~\cite{Firestone1991}. The presence of a~second $0^+$ state at such low excitation energy is somewhat surprising when comparing with the systematics of excited states in neighboring nuclei. On the other hand, such a~low-lying $0_2^+$ state could indicate the presence of shape coexistence near the ground state of $^{140}$Sm. The spin assignments for the two states in question were clarified in a~recent measurement of $\gamma$-$\gamma$ angular correlations following the $\beta$ decay of $^{140}$Eu, which firmly showed that the state at 990~keV in $^{140}$Sm has spin-parity $2^+$ and the state at 1599~keV $0^{(+)}$~\cite{Samorajczyk2015}. The level scheme for the states of the ground-state band and the lowest non-yrast states is shown in Fig.~\ref{fig:level}.

The collectivity of the $2_1^+$ state in $^{140}$Sm was studied in a~recent recoil distance lifetime measurement~\cite{Bello2015}. The resulting $B(E2;2_1^+ \rightarrow 0_1^+)$ value and its comparison to the neigboring samarium isotopes indicates a~gradual onset of deformation when removing neutrons from the closed-shell nucleus $^{144}$Sm$_{82}$. The experimental $B(E2;2_1^+ \rightarrow 0_1^+)$ values for the entire chain of neutron-deficient samarium isotopes are well reproduced by microscopic HFB calculations with the Gogny D1S interaction with mapping to the five-dimensional collective Hamiltonian (5DCH) for quadrupole excitations~\cite{Bello2015}. The theoretical calculations explain the onset of quadrupole collectivity in the samarium isotopes below $N=82$ by a~gradual shape transition from large prolate deformation with axial symmetry for the lightest isotopes to spherical shape in $^{144}$Sm, with the triaxial degree of freedom becoming more important as $N=82$ is approached. The Gogny 5DCH calculations find average quadrupole parameters of $\langle \beta \rangle = 0.17$ and $\langle \gamma \rangle = 29^{\circ}$ for the ground state of $^{140}$Sm~\cite{Bello2015}.

It appears that the onset of deformation in the samarium isotopes is different above and below the neutron shell closure. For $N>82$ the samarium nuclei show an abrupt increase in deformation at $N=90$, whereas the onset of deformation is more gradual for $N<82$ with apparent triaxiality in the transitional region around $^{140}$Sm. The rapid increase in deformation from $N=88$ to $N=90$ in the samarium and gadolinium isotopes can be explained by the occupation of proton and neutron spin-orbit partner orbitals which causes the disappearance of the $Z=64$ 
sub-shell closure for $N \geq 90$~\cite{Casten1981}. The transitional nucleus $^{152}$Sm was identified as one of the best realizations of the so-called X(5) critical point symmetry~\cite{Iachello2001,Casten2001PRL87}, which represents a~first-order phase transition from a~spherical vibrational to a~deformed rotational nucleus. The analysis of excitation spectra for the samarium isotopes with $N<82$ using the interacting boson approximation (IBA) found model parameters for $^{138}$Sm and $^{140}$Sm that place these nuclei between the U(5) and the SO(6) limits of the IBA~\cite{Pascu2010}, i.e.\ at the transition between a~spherical vibrator and a~$\gamma$-soft vibrator characterized by the so-called E(5) critical point symmetry~\cite{Iachello2000}. However, without any experimental $B(E2)$ values beyond the $2_1^+$ state and uncertain spin assignments for the lowest non-yrast states it was not possible to evaluate a~possible E(5) character for $^{140}$Sm. 

In this article we present the results of a~Coulomb excitation experiment using a~radioactive $^{140}$Sm beam, which provides several $B(E2)$ values and also the spectroscopic quadrupole moment for the $2_1^+$ state. Low-energy Coulomb excitation is an ideal method to study the collectivity and deformation in this nucleus, since both yrast and non-yrast states can be populated and transition probabilities extracted without interference from the isomeric $10^+$ states. The comparison of the experimental results with theoretical calculations sheds more light on the onset of deformation and collectivity in this mass region. The article is organized as follows: The experimental details and data analysis are described in Sec.~\ref{sec:experiment}. The extraction of electromagnetic matrix elements from the Coulomb excitation yields is described in Sec.~\ref{sec:coulomb_excitation}. The results are discussed and compared to theoretical calculations in Sec.~\ref{sec:discussion}, followed by a~summary and conclusions in Sec.~\ref{sec:summary}.

\section{Experimental details}\label{sec:experiment}

The application of the isotope separation on-line (ISOL) technique in combination with selective laser ionization and post-acceleration made it possible to study the electromagnetic properties of $^{140}$Sm by Coulomb excitation. Radioactive $^{140}$Sm atoms with a~half-life of 14.8 min were produced at the CERN-ISOLDE facility by spallation of a~primary tantalum target with 
$1.4$~GeV protons from the PS Booster. Samarium atoms were selectively ionized using the Resonance Ionization Laser Ion Source (RILIS)~\cite{Fedosseev2012}, which was equipped with a~
GdB${_6}$ low-work function cavity~\cite{Schwellnus2009} to reduce the surface ionization of isobaric impurities. 
After selection of mass $A=140$ using the general purpose separator GPS, the ions were bunched, cooled, and trapped in the REXTRAP \cite{Schmidt2002}, and then further ionized to charge state $34^+$ using the EBIS charge breeder \cite{Wenander2002}. Finally the highly charged $^{140}$Sm ions were accelerated to an energy of $2.85 A$~MeV in the REX linear accelerator \cite{Kester2003}. An average intensity of $2\times 10^5$ particles per second was achieved over a~beam time of approximately 100 hours. 

The $^{140}$Sm projectiles were scattered on a~secondary ${}^{94}$Mo target of 
$2$~mg/cm${}^2$ thickness. Both $^{140}$Sm projectiles and $^{94}$Mo target nuclei were excited in the low-energy Coulomb excitation reaction. The distance of closest approach of $19.2$~fm between the projectiles and target nuclei for the given reaction parameters is larger than the distance of $17.2$~fm, which is obtained by applying the safe distance separation criterion \cite{Cline1986}
\begin{equation}
	d > 1.25(A_p^{1/3}+A_t^{1/3}) + 5 {}\text{ fm}\label{eq:xdef3},
\end{equation}
where $A_p$ and $A_t$ are the mass numbers of the projectile and target, respectively. Under these conditions, it is safe to neglect all influence of the strong nuclear force and assume that the excitation process can be described by a~pure electromagnetic interaction. 

Gamma rays from excited states in the $^{140}$Sm projectiles and $^{94}$Mo target nuclei were detected in the MINIBALL HPGe detector array~\cite{Warr2013}, which at the time of the experiment consisted of seven triple-cluster modules, each of which comprised three six-fold segmented germanium crystals. An energy and efficiency calibration for the germanium detectors was performed using standard $^{152}$Eu and $^{133}$Ba sources. An annular double-sided silicon strip detector (DSSSD) of $1000$~$\mu$m thickness was mounted in the MINIBALL target chamber and used to detect both scattered projectiles and recoiling target nuclei. The DSSSD consisted of four individual quadrants with each $16$ concentric annular strips on the front side and $12$ azimuthal sector strips on the back side. The annular strips had a~strip pitch of $2$~mm and the azimuthal strips covered $3.5^{\circ}$ each. In total, the DSSSD covered $5000$~mm$^2$ with an active area of $93\%$~\cite{Ostrowski2002}. The detector was mounted in forward direction at a~distance of $25.2$~mm from the target, covering the angular range from $19.7^{\circ}$ to $58.4^{\circ}$ in the laboratory frame. In the center-of-mass frame, the angular coverage corresponded to $49.7^{\circ} < \theta_{CM} < 146^{\circ}$ for the detection of $^{140}$Sm projectiles and $63.1^{\circ} < \theta_{CM} < 140.7^{\circ}$ for the detection of recoiling target nuclei. 

\begin{figure}[b]
	\centering
	\includegraphics[width=\columnwidth]{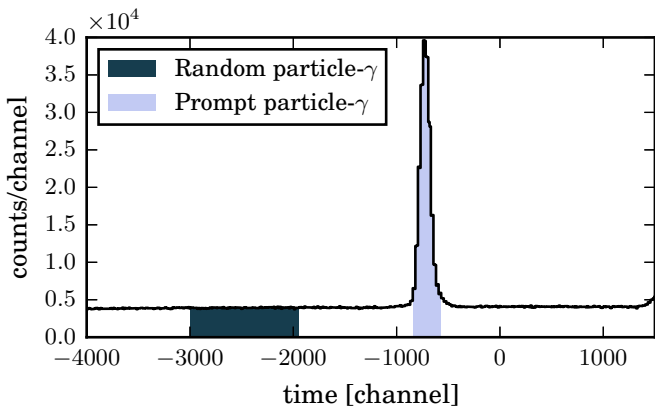}
	\caption{Particle-$\gamma$ coincidence time spectrum. A normalized fraction of the random gate was used for background subtraction.}
	\label{fig:time_gate}
\end{figure}

\begin{figure}[t]
	\centering
    \includegraphics[width=\columnwidth]{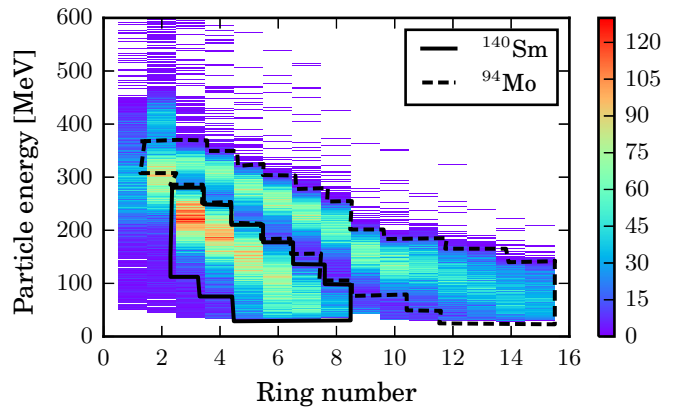}
	\caption{Energy spectra of particles detected in the DSSSD as a~function of the laboratory scattering angle. The cuts to select between detected $^{140}$Sm projectiles and recoiling $^{94}$Mo target nuclei are marked. Ring number $1$ corresponds to lab angle $19.7^\circ$ and ring $16$ to lab angle $58.4^\circ$.}
	\label{fig:dsssd}
\end{figure}

The data acquisition was triggered and events were built when the MINIBALL and DSSSD detectors gave coincident signals. Prompt particle-$\gamma$ coincidences were selected by applying a~prompt gate in the time spectrum, and background from random coincidences was subtracted as shown in Fig.~\ref{fig:time_gate}. 
The energy spectrum for particles as a~function of scattering angle measured with one DSSSD quadrant in coincidence with $\gamma$-rays detected in MINIBALL is shown in Fig.~\ref{fig:dsssd}. The contours illustrate the separation and selection for scattered projectiles and recoiling target nuclei. The innermost annular strips of the DSSSD could not be used in the further analysis because of insufficient separation between the projectiles and target nuclei. The inverse kinematics with heavier projectiles and lighter target nuclei leads to an ambiguity where different center-of-mass scattering angles result in the same laboratory detection angle for the $^{140}$Sm projectiles, which is further complicated by the low-energy cut-off for projectiles with the largest center-of-mass scattering angles. For this reason, the detection angle of the  recoiling target nuclei was used to determine the center-of-mass scattering angle. For those center-of-mass scattering angles where both projectiles and recoiling target nuclei are expected to hit the sensitive region of the DSSSD, it was required that both particles were detected in opposite quadrants of the detector. The reaction kinematics and the hit pattern in the DSSSD were furthermore used to verify that the beam was well centered.

The $\gamma$ rays detected in MINIBALL are Doppler shifted depending on the velocity and angle with respect to the emitting particle. Since it is not possible to determine if a given $\gamma$ ray is emitted from a $^{140}$Sm or $^{94}$Mo nucleus, two spectra are produced with the assumption that all $\gamma$ rays are emitted by either the projectiles or target nuclei and by applying the appropriate Doppler correction using the information from the DSSSD. This procedure results in two spectra where some transitions were properly Doppler corrected and appear as sharp peaks, whereas others are very broad due to a~wrong Doppler correction. As long as the sharp and broad peaks are not overlapping it is possible to determine the intensities of the transitions from the respective spectra. 

\begin{figure}[t]
	\centering
    \includegraphics[width=\columnwidth]{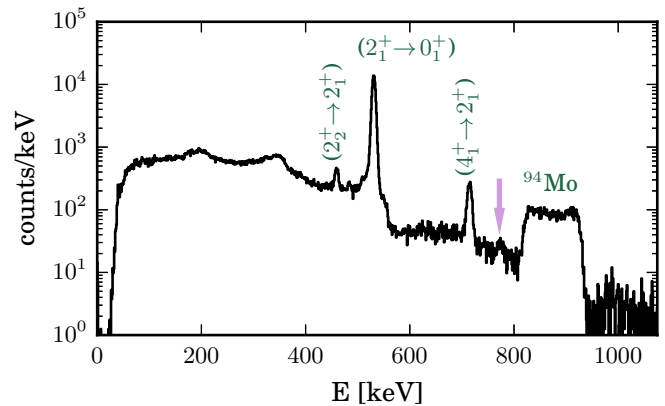}
	\caption{Background subtracted $\gamma$-ray spectrum in coincidence with a particle in the DSSSD, with Doppler correction for ${}^{140}$Sm. The $2_1^+\rightarrow 0_1^+$ transition at 531~keV, the $4_1^+\rightarrow 2_1^+$ transition at 715~keV, and the $2_2^+\rightarrow 2_1^+$ transition at 460~keV are visible. The broad structure originates from the 871~keV $2_1^+\rightarrow 0_1^+$ transition in $^{94}$Mo. The purple arrow marks the $2_1 ^+\rightarrow0_1^+$ transition of the potential beam contaminant ${}^{140}$Nd.}
	\label{fig:140Sm_spec}
\end{figure}

\begin{figure}[t]
	\centering
    \includegraphics[width=\columnwidth]{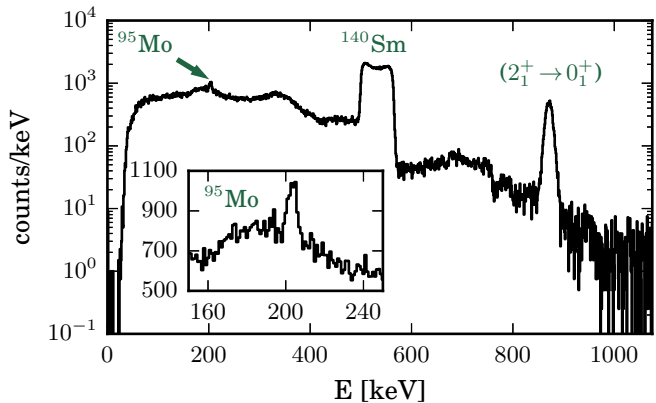}
	\caption{Background subtracted $\gamma$-ray spectrum in coincidence with a particle in the DSSSD, with Doppler correction for the ${}^{94}$Mo recoils. The $2_1^+\rightarrow 0_1^+$ transition in ${}^{94}$Mo was observed at 871~keV. The inset shows an enlarged part of the same spectrum, where the $3/2^+ \rightarrow 5/2^+_{\rm gs}$ transition in the target contaminant ${}^{95}$Mo is seen at 204~keV.}
	\label{fig:Mo_spec}
\end{figure}

Background subtracted spectra with Doppler correction for ${}^{140}$Sm and ${}^{94}$Mo are shown in Figs.~\ref{fig:140Sm_spec} and~\ref{fig:Mo_spec}, respectively. The $2_1^+\rightarrow 0_1^+$ transition at 531 keV, the $4_1^+\rightarrow 2_1^+$ transition at 715 keV, and the $2_2^+\rightarrow 2_1^+$ transition at 460 keV are visible in the spectrum that is Doppler corrected for the ${}^{140}$Sm velocity. A hint of a~transition is visible at 774 keV, marked by a purple arrow in Fig.~\ref{fig:140Sm_spec}. The energy corresponds to 
the $2_1^+ \rightarrow 0_1^+$ transition in $^{140}$Nd, which could be present as a~contaminant in the beam. From the intensity of the transition this contamination is estimated to have an upper limit of $0.8\%$ of the total beam intensity. The $\gamma$-ray energy spectrum with Doppler correction for the projectiles contains no indication of other beam contaminants. 

\begin{figure}[b]
	\centering
    \includegraphics[width=\columnwidth]{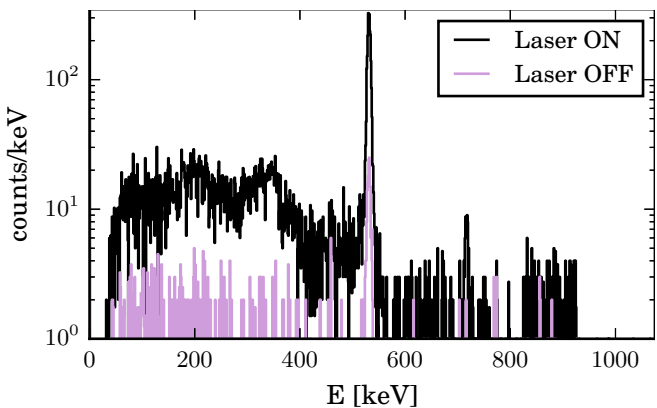}
	\caption{Gamma-ray spectra obtained with and without RILIS laser ionization in a~measurement during which the lasers were periodically switched on and off. The spectra are Doppler corrected for the $^{140}$Sm projectiles.}
	\label{fig:laser}
\end{figure}

The beam composition was monitored by performing regular measurements during which the RILIS lasers were periodically switched on and off. Fig.~\ref{fig:laser} shows the spectra acquired during periods in which the lasers were turned on and off, respectively. Even in the spectrum with the lasers turned off only the $2_1^+ \rightarrow 0_1^+$ transition in $^{140}$Sm is visible, although with much lower intensity compared to the spectrum taken with laser ionization. The acceleration of $^{140}$Sm without laser is due to surface ionization. The presence of any beam contaminants should be enhanced in the spectrum taken without lasers. The absence of any other transitions further supports the conclusion that the beam was composed of at least $99.2$\% $^{140}$Sm during the measurements with the lasers switched on.

An additional germanium detector was placed behind the beam dump downstream from MINIBALL to monitor the $\gamma$-ray spectra following the $\beta$ decay of the beam particles. The spectra taken with this detector show no evidence for any radioactive beam contaminants. The small possible $^{140}$Nd contaminant was considered negligible, and all further analysis was performed under the assumption that the $^{140}$Sm beam was pure. 

Fig.~\ref{fig:Mo_spec} shows the total $\gamma$-ray spectrum with Doppler correction for the recoiling $^{94}$Mo target nuclei. The $2_1^+ \rightarrow 0_1^+$ transition in $^{94}$Mo at 871 keV is clearly visible, whereas the transitions in $^{140}$Sm appear as broad structures. Closer inspection reveals a~weak line at 204 keV. This peak is most likely from the $3/2^+ \rightarrow 5/2^+_{\rm gs}$ transition in $^{95}$Mo, which could be present as a small isotopic contamination in the target. This interpretation is supported by the fact that the same weak transition was also observed in previous experiments using the same target foil \cite{Kesteloot2015}. Since the electromagnetic matrix elements are well known for both $^{94}$Mo \cite{Abriola2006} and $^{95}$Mo \cite{Basu2010}, the amount of $^{95}$Mo in the target can be determined from the Coulomb excitation analysis, as will be discussed below.

\section{Coulomb excitation data analysis}\label{sec:coulomb_excitation}

\begin{figure*}
	\centering
    \includegraphics[width=\textwidth]{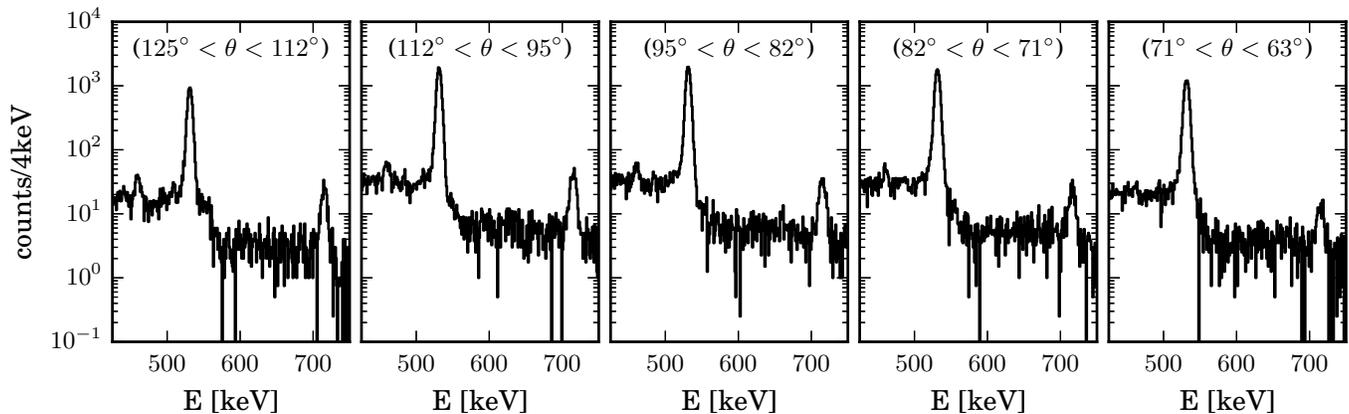}
	\caption{Gamma-ray spectra for five separate ranges of center-of-mass scattering angles with Doppler correction for $\gamma$ emission from the 
$^{140}$Sm projectiles.}
	\label{fig:5_spec}
\end{figure*}

The analysis of the Coulomb excitation data and extraction of electromagnetic matrix elements utilizes the angular dependence of the differential Coulomb excitation cross sections. For this purpose, the data were subdivided into various ranges of scattering angles as measured with the DSSSD. It was found that a division into five angular bins represents a good compromise between the maximum number of data points for differential cross sections and the minimum level of statistics in the spectra corresponding to each angular bin. The resulting spectra for five angular ranges and with Doppler correction for $\gamma$-ray emission from the $^{140}$Sm projectiles are shown in Fig.~\ref{fig:5_spec}. The spectra reveal how the relative strengths of the $4_1^+ \rightarrow 2_1^+$ and $2_2^+ \rightarrow 2_1^+$ transitions, which require two-step excitations, change with scattering angle compared to the $2_1^+ \rightarrow 0_1^+$ transition. The measured intensities for the three observed transitions in $^{140}$Sm, the $2_1^+ \rightarrow 0_1^+$ transition in $^{94}$Mo, and the $3/2_1^+ \rightarrow 5/2_{gs}^+$ transition in the target contaminant $^{95}$Mo are listed in Table \ref{table:gamma_int5} for the five angular ranges. 

\begin{table*}[htb]
\caption{Measured $\gamma$-ray intensities for all observed transitions for the five ranges of center-of-mass scattering angles. Values are rounded to two significant figures in the uncertainty.}
\centering
\begin{tabular*}{15 cm}{@{\extracolsep{\fill}}cc D{,}{\pm}{-1} D{,}{\pm}{-1} D{,}{\pm}{-1} D{,}{\pm}{-1} D{,}{\pm}{-1}}
\hline\hline
\multicolumn{2}{c}{transition} & 
\multicolumn{1}{c}{[63$^{\circ}$, 71$^{\circ}$]} &
\multicolumn{1}{c}{[71$^{\circ}$, 82$^{\circ}$]} &
\multicolumn{1}{c}{[82$^{\circ}$, 95$^{\circ}$]} &
\multicolumn{1}{c}{[95$^{\circ}$, 112$^{\circ}$]} &
\multicolumn{1}{c}{[112$^{\circ}$, 125$^{\circ}$]} \\
\hline
$^{140}$Sm & $2_1^+ \rightarrow 0_1^+$ & 
10200,140 & 13840,170 & 14350,170 & 13780,170 & 6860,120 \\
$^{140}$Sm & $4_1^+ \rightarrow 2_1^+$ & 
118,18 & 209,24 & 286,31 & 389,34 & 199,23 \\
$^{140}$Sm & $2_2^+ \rightarrow 2_1^+$ & 
47,22 & 150,33 & 239,58 & 306,60 & 300,110 \\
$^{94}$Mo & $2_1^+ \rightarrow 0_1^+$ & 
743,39 & 1073,47 & 1213,50 & 1090,47 & 543,34 \\
$^{95}$Mo & $3/2_1^+ \rightarrow 5/2_{gs}^+$ &
175,40 & 197,45 & 249,43 & 224,53 & 179,62 \\ 
\hline \hline
\end{tabular*}
\label{table:gamma_int5}
\end{table*}

The coupled channel code GOSIA \cite{Czosnyka1983,Cline2011} was used to extract the electromagnetic matrix elements. The program combines semi-classical Coulomb excitation calculations with a multidimensional fitting procedure of the matrix elements. In this procedure, the set of matrix elements is found that best reproduces the measured $\gamma$-ray intensities observed for the different ranges of scattering angles, taking into account the geometry and efficiency of both particle and $\gamma$-ray detectors. Known lifetimes, branching ratios, and mixing ratios can be included in the $\chi^2$ minimization. The $\gamma$-ray yields are obtained by integrating the Coulomb excitation cross section over the range of scattering angles covered by the experiment and integrating over the range of projectile energies resulting from the energy loss in the target. The measured $\gamma$-ray intensities were corrected for the relative efficiency values obtained from source calibrations. Finally, the correlated uncertainties were calculated for the set of reduced transitional and diagonal matrix elements. 

To convert the measured $\gamma$-ray intensities into absolute cross sections, it is possible to normalize to the elastic Rutherford cross section obtained from particle-singles events. However, this requires precise knowledge of the particle detector efficiency, dead time, and beam intensity. The latter is often difficult to obtain with good precision in experiments with weak radioactive ion beams. In cases where the lifetime of one or more excited states are known, the corresponding reduced matrix element can be used to obtain absolute cross sections in the normalization procedure. Without prior knowledge of matrix elements, a different normalization technique is required. 
The GOSIA2 code \cite{Zielinska2015,Cline2011} was developed to allow for the simultaneous analysis of both projectile and target excitation, using known reduced matrix elements for the scattering partners in the normalization. The ratio of observed transitions from the projectiles $N_p$ and the target $N_t$ is independent of the particle detection efficiency $\epsilon_{part}$ and the time-integrated luminosity $L$, as seen in Eq.~\eqref{eq:norm}:
\begin{equation}
	\frac{N_p}{N_t}= \frac{L\epsilon_{part}b_p\epsilon_{\gamma}(E_p)\sigma_p}{L\epsilon_{part}b_t\epsilon_{\gamma}(E_t)\sigma_t},
\label{eq:norm}
\end{equation}
where $b_p$ and $b_t$ are the $\gamma$-ray branching ratios, $\epsilon_{\gamma}(E_p)$ and $\epsilon_{\gamma}(E_t)$ the $\gamma$-ray efficiencies, and $\sigma_p$ and $\sigma_t$ the integrated cross sections for projectile and target excitation, respectively. 

Two approaches were used in the data analysis from the current $^{140}$Sm Coulomb excitation experiment. Since the low-spin structure of $^{140}$Sm was initially unknown, the first approach was based on the normalization to the target excitation. After the measurement of the lifetime of the $2_1^+$ state in $^{140}$Sm \cite{Bello2015} it was also possible to use the resulting $B(E2)$ value for normalization. A detailed description of both analysis approaches is presented below.

\subsection{Normalization to the target excitation}
\label{sec:coulex:target}

Several criteria had to be considered for the choice of the target material in the Coulomb excitation experiment. Since the energy of the REX post-accelerator is limited to 3~$A$ MeV, high-$Z$ materials are disadvantageous because they would lead to large distances between the scattering partners and consequently to low cross sections. The mass of the target nucleus also has to be sufficiently different from the mass of the projectile to avoid ambiguities in the kinematics and ensure sufficient separation between the projectile and target nuclei in the plot of the particle energy as a function of scattering angle ({\it c.f.}~Fig.~\ref{fig:dsssd}). The energies of the $\gamma$-ray transitions in the projectile and target nuclei should be well separated to avoid overlapping peaks in the spectra. Finally, the matrix elements for the low-lying states should be well known to use the excitation of the target nucleus as normalization. It was found that $^{94}$Mo was a suitable target material fulfilling the above criteria. The relevant electromagnetic matrix elements are known from Coulomb excitation experiments with $\alpha$ and $^{16}$O projectiles \cite{Barrette1972,Paradis1976}. 

As mentioned above, the observation of a transition at 204 keV in the spectrum that was Doppler corrected for the target recoils suggests the presence of $^{95}$Mo in the target foil. The number of $^{95}$Mo atoms, $N_{95}$, relative to the number of $^{94}$Mo atoms, $N_{94}$, in the target foil can be found as:
\begin{equation}	
\frac{N_{95}}{N_{94}} =
\frac{Y_{95}}{Y_{94}} \frac{\epsilon_{95}}{\epsilon_{94}}
\frac{\sigma_{94}}{\sigma_{95}}
\label{eq:95cont}
\end{equation}
where $Y_{A}$ and $\epsilon_A$ are the respective $\gamma$-ray yields and efficiencies for the $2_1^+\rightarrow 0_1^+$ and $3/2^+\rightarrow 5/2^+$ transitions in the two isotopes, and the cross sections $\sigma_A$ for the populations of the $2_1^+$ and $3/2^+$ states that are calculated from the known reduced matrix elements. An admixture of 4.4(11)\% $^{95}$Mo is found in this way, which is in good agreement with the value of 5(2)\% that was found in another Coulomb excitation experiment using the same target foil \cite{Kesteloot2015}.

\begin{figure*}
\centering
\includegraphics[width=\textwidth]{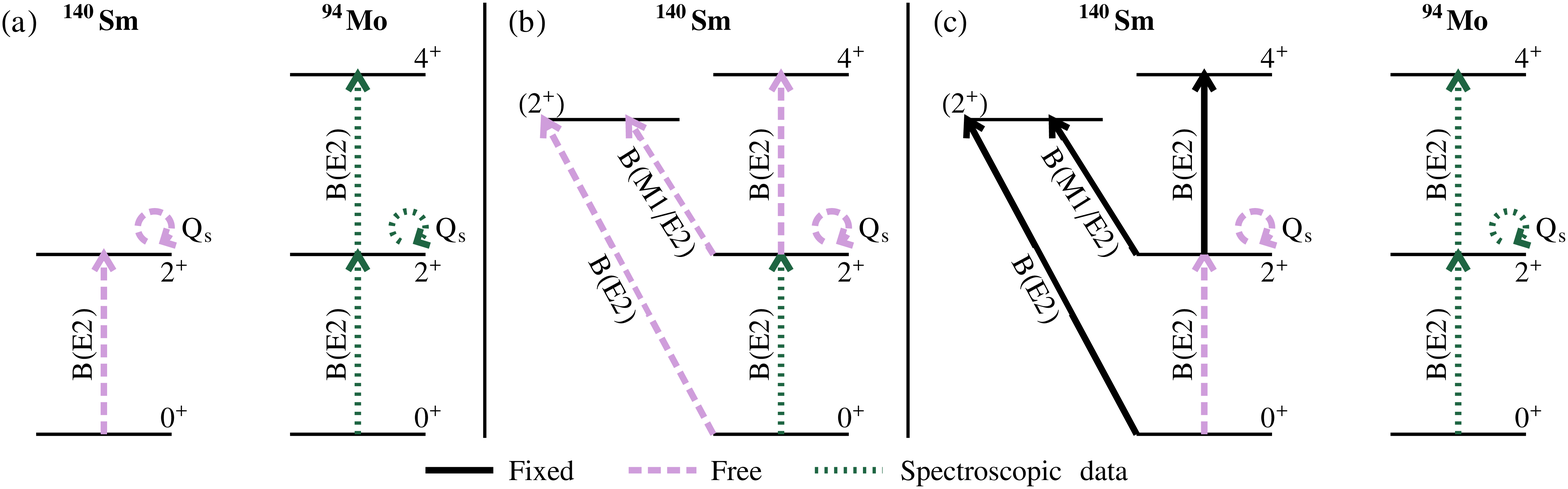}
\caption{Illustration of the iterative procedure to fit the reduced matrix elements 
using the codes GOSIA and GOSIA2, with part a, b and c corresponding to steps 2, 3, and 4 as described in the text. The level schemes indicate which matrix elements were included in the fit as free parameters, as fixed values, or as free parameters with spectroscopic data to constrain the fit. Iterations between step 3 and step 4 were performed until the solution converged.}
\label{fig:LevelScheme}
\end{figure*}

An iterative procedure with alternating use of the codes GOSIA and GOSIA2 was employed to determine reduced matrix elements in $^{140}$Sm with normalization to the target excitation. The procedure, which is explained in more detail elsewhere \cite{Zielinska2015}, is illustrated in Fig.~\ref{fig:LevelScheme} and summarized below.

\begin{enumerate}
	\item In the first step, the standard GOSIA code is used for $^{94}$Mo with the intensities of the $2_1^+ \rightarrow 0_1^+$ transition as input data and the known reduced matrix elements included as spectroscopic data. In this way an initial set of normalization factors was obtained.

	\item In the second step, the code GOSIA2 was used to simultaneously fit the reduced matrix elements in both $^{140}$Sm and $^{94}$Mo. The purpose of this step was to obtain a first estimate of the $\langle 0_1^+  \| E2 \| 2_1^+ \rangle$ matrix element in $^{140}$Sm. Therefore only the matrix elements $\langle 0_1^+  \| E2 \| 2_1^+ \rangle$ and  $\langle 2_1^+  \| E2 \| 2_1^+ \rangle$ and the $\gamma$-ray intensities for the $2_1^+ \rightarrow 0_1^+$ transition were included for $^{140}$Sm ({\it c.f.}~Fig.~\ref{fig:LevelScheme}a). To minimize the influence of multistep Coulomb excitation, only data for $\theta_{CM} < 100^{\circ}$ were included in this step. With the high level of statistics for the $2_1^+ \rightarrow 0_1^+$ transition, the data could be divided into ten angular ranges for this step, resulting in a better sensitivity for the $\langle 2_1^+  \| E2 \| 2_1^+ \rangle$ matrix element. The measured intensities for this subdivision of the data are given in Table \ref{table:gamma_10}. With the normalization from step 1 as starting values, the $\chi^2$ minimization yielded a new set of normalization factors and the $\langle 0_1^+  \| E2 \| 2_1^+ \rangle$ and $\langle 2_1^+  \| E2 \| 2_1^+ \rangle$ matrix elements for $^{140}$Sm. Fig.~\ref{fig:chi} shows the two-dimensional $\chi^2$ map for the two samarium matrix elements. Their uncertainties were obtained from the contour for $\chi^2_{\min} + 1$. 

\begin{table}[b]
\caption{Counts observed for the $2_1^+ \rightarrow 0_1^+$ transitions in the $^{140}$Sm projectiles and the $^{94}$Mo target nuclei and their uncertainties for ten different ranges of scattering angles in the center-of-mass frame. These are the values used in the second step of the GOSIA-GOSIA2 iteration where $\theta_{CM} < 100^{\circ}$. Values are rounded to two significant figures in the uncertainty.}
\centering
\begin{tabular*}{7 cm}{@{\extracolsep{\fill}}c D{,}{\pm}{-1} D{,}{\pm}{-1}}
\hline \hline
angular range&\multicolumn{1}{c}{$^{140}$Sm}&\multicolumn{1}{c}{$^{94}$Mo}\\
\hline
{[}95$^{\circ}$, 100$^{\circ}$]& 4770,240 & 397,29 \\
{[}90$^{\circ}$, 95$^{\circ}$] & 3290,160 & 245,23 \\
{[}86$^{\circ}$, 90$^{\circ}$] & 4300,220 & 351,27 \\
{[}82$^{\circ}$, 86$^{\circ}$] & 5040,250 & 450,31 \\
{[}78$^{\circ}$, 82$^{\circ}$] & 5030,250 & 439,30 \\
{[}75$^{\circ}$, 78$^{\circ}$] & 4320,230 & 311,26 \\
{[}71$^{\circ}$, 75$^{\circ}$] & 4010,200 & 323,26 \\
{[}68$^{\circ}$, 71$^{\circ}$] & 4070,200 & 310,26 \\
{[}66$^{\circ}$, 68$^{\circ}$] & 3560,180 & 250,23 \\
{[}63$^{\circ}$, 66$^{\circ}$] & 2610,130 & 206,24 \\
\hline \hline
\end{tabular*}
\label{table:gamma_10}
\end{table}

\begin{figure}[b]
\centering
\includegraphics[width=\columnwidth]{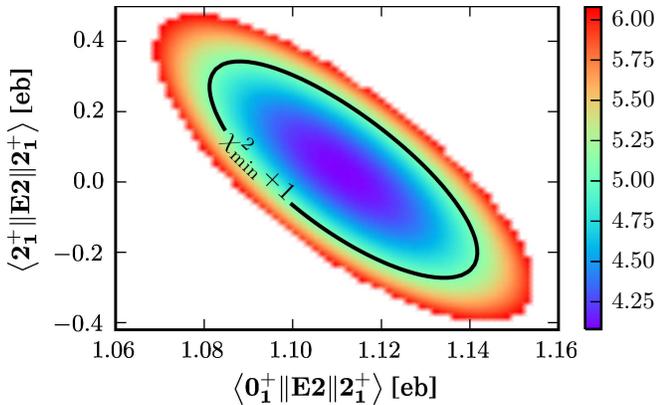}
\caption{Result of the $\chi^2$ minimization for the $\langle 0_1^+ \| E2 \| 2_1^+ \rangle$ and $\langle 2_1^+ \| E2 \| 2_1^+ \rangle$ matrix elements in $^{140}$Sm obtained during step 2 of the iterative procedure using target normalization.}
\label{fig:chi}
\end{figure}

	\item In the third step, the standard GOSIA code was used for $^{140}$Sm. 
The $\langle 0_1^+  \| E2 \| 2_1^+ \rangle$ matrix element obtained in step 2 and its uncertainty were treated as known spectroscopic data, whereas all other relevant matrix elements, $\langle 2_1^+  \| E2 \| 2_1^+ \rangle$, $\langle 2_1^+  \| E2 \| 4_1^+ \rangle$, $\langle 2_1^+  \| E2 \| 2_2^+ \rangle$, $\langle 2_1^+  \| M1 \| 2_2^+ \rangle$, and $\langle 0_1^+  \| E2 \| 2_2^+ \rangle$, were treated as free parameters 
({\it c.f.}~Fig.~\ref{fig:LevelScheme}b). In this step the $\gamma$-ray intensities from the division into five bins covering the entire angular range ({\it c.f.}~Table \ref{table:gamma_int5}) were used to ensure sufficient statistics for the $4_1^+ \rightarrow 2_1^+$ and $2_2^+ \rightarrow 2_1^+$ transitions. An upper limit was included for the unobserved $2_2^+ \rightarrow 0_1^+$ transition. This step yielded a first realistic estimate of all relevant reduced matrix elements in $^{140}$Sm. 

	\item In the fourth step, the code GOSIA2 was again used for simultaneous $\chi^2$ minimization of reduced matrix elements in $^{140}$Sm and $^{94}$Mo. Only the $\langle 0_1^+ \| E2 \| 2_1^+ \rangle$ and $\langle 2_1^+ \| E2 \| 2_1^+ \rangle$ matrix elements in $^{140}$Sm were treated as free parameters; all other matrix elements for $^{140}$Sm were fixed to the values from the previous step
({\it c.f.}~Fig.~\ref{fig:LevelScheme}c). The $\gamma$-ray intensities from the division into five angular ranges were included for all observed transitions. This step yielded more realistic values for $\langle 0_1^+ \| E2 \| 2_1^+ \rangle$ and $\langle 2_1^+ \| E2 \| 2_1^+ \rangle$, since effects from coupling to higher-lying states were taken into account. Uncertainties for the free matrix elements were again taken from the $\chi^2_{\min} + 1$ contour in the $\chi^2$ map. In addition, a new set of normalization factors was obtained.
\end{enumerate}

Steps three and four of the GOSIA-GOSIA2 iteration procedure were then repeated until the final solution stabilized. Overall uncertainties of all determined matrix elements were calculated by using the GOSIA code as the last stage of the data analysis. The resulting reduced matrix elements are presented in Table \ref{table:result}. Reduced transition probabilities and spectroscopic quadrupole moments can be extracted from the transitional and diagonal matrix elements, respectively, using the following relations:
\begin{equation}
B(E2; I_i\rightarrow I_f) = 
\frac{\arrowvert\langle I_f\arrowvert\arrowvert E2 \arrowvert\arrowvert I_i\rangle\arrowvert^2}{2I_i+1},
\label{eq:xdef4}
\end{equation}
\begin{equation}
Q_s(I) =\sqrt{\frac{16\pi}{5}}\frac{\langle II20 \arrowvert II\rangle}{\sqrt{2I+1}}{\langle I \| E2 \| I\rangle},
\label{eq:xdef5}
\end{equation}
where ${\langle II20 \arrowvert II\rangle}$ is a Clebsch-Gordan coefficient.
The data, and by consequence also the $\chi^2$ minimization, was very insensitive to the $\langle 2_2^+ \|M1|\ 2_1^+ \rangle$ matrix element, and no reliable value could be extracted. The $\langle 2_2^+ \| E2 \| 0_1^+ \rangle$ matrix element was also included in the analysis together with an upper limit for the intensity of the transition, which yielded an upper limit for the  transition strength, $B(E2;2_2^+ \rightarrow 0_1^+) < 0.001$ $e^2$b$^2$. 

\begin{table}[h]
\caption{Reduced matrix elements for ${}^{140}$Sm and associated $B(E2)$ values and spectroscopic quadrupole moment with correlated errors obtained with the target normalization approach.}
\centering
\begin{tabular*}{\columnwidth}{@{\extracolsep{\fill}}cccc}
\hline \hline
 $I_i$ & $I_f$ 
 & ${\langle I_f \| E2 \| I_i\rangle}$ [$e$b]
 & $B(E2; I_i\rightarrow I_f)$ [$e^2$b$^2$] \\
\hline
$2_1^+$ &$0_1^+$ &1.11$_{-0.03}^{+0.03}$ &0.25$_{-0.01}^{+0.02}$\\
$4_1^+$ &$2_1^+$ &1.63$_{-0.05}^{+0.05}$ &0.30$_{-0.02}^{+0.02}$\\
$2_2^+$ &$2_1^+$ &1.33$_{-0.09}^{+0.08}$ &0.35$_{-0.05}^{+0.05}$\\
\hline
\multicolumn{2}{c}{$I$} 
 & ${\langle I \arrowvert\arrowvert E2 \arrowvert\arrowvert I \rangle}$ [$e$b]
 & $Q_s(I)$ [$e$b]\\
  \hline
\multicolumn{2}{c}{$2_1^+$} & +0.03$_{-0.20}^{+0.54}$ & +0.02$_{-0.15}^{+0.41}$\\
\hline \hline
\end{tabular*}
\label{table:result}
\end{table}

\subsection{Normalization to the lifetime of the $2_1^+$ state}
\label{sec:coulex:lifetime}

The lifetime of the $2_1^+$ state in $^{140}$Sm was recently measured to be 9.1(6) ps in an experiment using the recoil-distance Doppler shift technique~\cite{Bello2015}. The resulting matrix element, $\langle 0_1^+\|E2\|2_1^+ \rangle = 1.03(3)$~$e$b, can be used to normalize the Coulomb excitation data instead of normalizing to the excitation of the $^{94}$Mo target nuclei. In this case, the lifetime is included as an additional data point in the $\chi^2$ minimization within the standard GOSIA code. The $\chi^2$ minimization is equivalent to step 3 in the iterative procedure described 
in section \ref{sec:coulex:target} and illustrated in Fig.~\ref{fig:LevelScheme}. The reduced matrix elements obtained in this way are presented in Table \ref{table:result_lt}.
Note that the obtained matrix element $\langle 0_1^+\|E2\|2_1^+ \rangle$ differ slightly from the value corresponding to the lifetime used as normalization. This is possible since the matrix element is allowed to vary in the $\chi^2$ minimization to best fit all available data.

\begin{table}[h]
\caption{Reduced matrix elements for ${}^{140}$Sm and associated $B(E2)$ values and spectroscopic quadrupole moment with correlated errors obtained with the lifetime normalization approach.}
\centering
\begin{tabular*}{\columnwidth}{@{\extracolsep{\fill}}cccc}
\hline \hline
 $I_i$ & $I_f$ 
& ${\langle I_f\arrowvert\arrowvert E2 \arrowvert\arrowvert I_i\rangle}$ [$e$b]
& $B(E2; I_i\rightarrow I_f)$ [$e^2$b$^2$] \\
\hline
$2_1^+$ &$0_1^+$ &$1.02_{-0.03}^{+0.04}$ &$0.21_{-0.01}^{+0.02}$\\
$4_1^+$ &$2_1^+$ &$1.61_{-0.05}^{+0.05}$ &$0.29_{-0.02}^{+0.02}$\\
$2_2^+$ &$2_1^+$ &$1.32_{-0.09}^{+0.08}$ &$0.35_{-0.05}^{+0.04}$\\
\hline
\multicolumn{2}{c}{$I$} 
& ${\langle I \|E2\| I \rangle}$ [$e$b] & $Q_s(I)$ [$e$b] \\
\hline
\multicolumn{2}{c}{$2_1^+$} & $-0.17_{-0.19}^{+0.51}$ &$-0.13_{-0.14}^{+0.38}$\\
\hline \hline
\end{tabular*}
\label{table:result_lt}
\end{table}

The resulting $B(E2;2_1^+ \rightarrow 0_1^+)$ value reproduces the value from the lifetime measurement and is slightly smaller than the value obtained using the target normalization approach. It is interesting to note that the matrix elements connecting the $2_1^+$ state with the $2_2^+$ and $4_1^+$ states are hardly affected and the values obtained with the lifetime normalization are almost identical to the ones obtained with target normalization. The quadrupole moment for the $2_1^+$ state becomes slightly more negative when normalizing to the lifetime. However, the value is still rather small and the two results agree well within their uncertainty.

The small discrepancy between the two normalization approaches for the $\langle 2_1^+ \|E2\| 0_1^+ \rangle$ matrix element could be due to unaccounted-for systematic errors. One possible source for such errors could be unknown target impurities. The discrepancy could also be due to systematic errors for the lifetime of the $2_1^+$ state in $^{140}$Sm or for the matrix elements in the $^{94}$Mo target nucleus.

\subsection{State at 990~keV}

In an earlier $\beta$ decay experiment the state at 990 keV excitation energy was tentatively assigned to have spin-parity $I^{\pi} = (0^+)$ \cite{Firestone1991}, but a more recent experiment revised this result and firmly assigned spin-parity $I^{\pi} = 2^+$ to this state \cite{Samorajczyk2015}. In the early stages of the analysis of the present Coulomb excitation experiment, the spin assignment for the state at 990 keV was not yet resolved, and the complete analysis, using both target and lifetime normalization approaches as described above, was performed with the assumption that the state at 990 keV had spin-parity $0^+$. The results concerning the $2_1^+$ and $4_1^+$ states were almost identical to the ones presented in Tables \ref{table:result} and \ref{table:result_lt}. However, the 
$\langle 0_2^+ \|E2\| 2_1^+ \rangle$ matrix element yielded $B(E2;0_2^+ \rightarrow 2_1^+) = 1.02(15)$ $e^2$b$^2$, which corresponds to 236 Weisskopf units. This very large transition probability was difficult to understand, and motivated the new experiment to measure the spin of the state at 990 keV using $\gamma-\gamma$ angular correlations. This experiment also found that the $2_2^+ \rightarrow 2_1^+$ transition is of almost pure $E2$ character with only a very small $M1$ admixture \cite{Samorajczyk2015}.

\section{Discussion}\label{sec:discussion}

\begin{figure*}
	\centering
    \includegraphics[width=\textwidth]{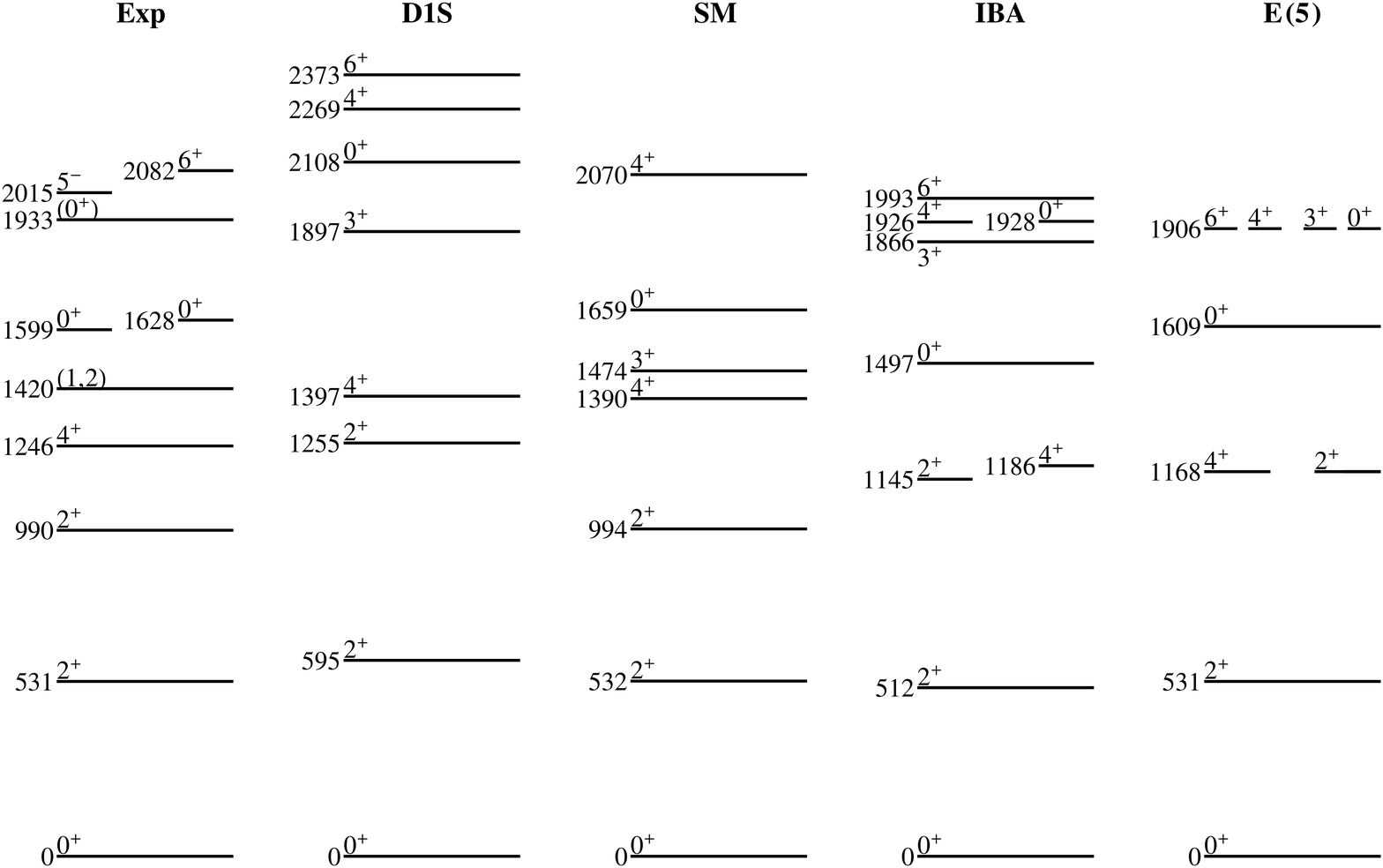}
	\caption{Comparison of the experimental excitation energies (in keV) of the low-lying states in $^{140}$Sm with predictions from beyond-mean field calculations based on the Gogny D1S interaction, the shell model (SM), the interacting boson approximation (IBA), and expectations for a nucleus with E(5) critical point symmetry (see text for explanations).
}
	\label{fig:Schemes}
\end{figure*}

The values using the target normalization approach are in good agreement with those obtained by normalizing to the measured lifetime of the $2_1^+$ state. Only for the $\langle 2_1^+ \| E2 \| 0_1^+ \rangle$ reduced matrix element is a discrepancy of $1.8 \sigma$ found between the two techniques. Both the target and lifetime normalization techniques rely on data from independent measurements. The former approach relies on reduced matrix elements for the $^{94}$Mo target nucleus measured in separate Coulomb excitation experiments \cite{Barrette1972,Paradis1976}, whereas the latter relies on the lifetime of the $2_1^+$ state in the $^{140}$Sm projectiles from a recoil-distance Doppler shift measurement \cite{Bello2015}. Without any obvious weaknesses in either approach or independent measurement, it is difficult to choose which results should be trusted more. We therefore adopt the average values from the two normalization methods for the following discussion. The adopted $B(E2)$ values are shown in Table \ref{table:theory} together with results from various theoretical calculations. The experimental excitation energies of the states are compared to the theoretical calculations in Fig.\ref{fig:Schemes}. 

\begin{table}[ht!]
\caption{Comparison of experimental and theoretical B$(E2;I_i\rightarrow I_f)$ [$e^2 b^2$] values and spectroscopic quadrupole moments, $Q_s$ [eb] for $^{140}$Sm.}
\centering
\begin{tabular}{llcccc}
\hline \hline
 $I_i$ & $I_f$ & Exp & D1S & SM & IBA\\
\hline
$2_1^+$ & $0_1^+$ & $0.23(2)$ & 0.208 & 0.218 & 0.219\\
$4_1^+$ & $2_1^+$ & $0.30(2)$ & 0.338 & 0.314 & 0.326\\
$6_1^+$ & $4_1^+$ &           & 0.455 &       & 0.379\\
$2_2^+$ & $2_1^+$ & $0.35(5)$ & 0.330 & 0.310 & 0.334\\
$2_2^+$ & $0_1^+$ & $<0.001$  & $2\times 10^{-5}$ & $1.9\times 10^{-4}$ & $2\times 10^{-4}$\\
$3_1^+$ & $2_1^+$ &           & $2\times 10^{-5}$ & $2.3\times 10^{-4}$ & $1\times 10^{-4}$\\
$3_1^+$ & $2_2^+$ &           & 0.396 & 0.362 & 0.277\\
$0_2^+$ & $2_1^+$ &           & 0.138 & 0.008 & 0.126\\
$0_2^+$ & $2_2^+$ &           & $2.6\times 10^{-3}$ & 0.241 & 0.114\\
 \hline
 & &$Q_s[eb]$& & &   \\
 \hline
$2_1^+$ &$2_1^+$&$-0.06_{-0.15}^{+0.41}$& -0.12 & -0.106 & -0.106 \\
$2_2^+$ &$2_2^+$&                    &  0.11 &  0.113 &  0.033 \\
\hline \hline
\end{tabular}
\label{table:theory}
\end{table}

\subsection{Geometric models}

The spectroscopic quadrupole moment for the $2_1^+$ state is consistent with zero, although the uncertainty is large. The spherical shape is inconsistent with an interpretation of the ground-state band as a rotational excitation with axial symmetry. An average spherical shape could indicate a quadrupole vibrational nature of the $2_1^+$ state. However, the $B(E2;4_1^+ \rightarrow 2_1^+)$ value is only insignificantly larger than the $B(E2;2_1^+ \rightarrow 0_1^+)$ value, and not twice as large as would be required for a~harmonic vibration. The energy ratio $E(4_1^+)/E(2_1^+)=2.35$ is typical for a~transitional nucleus between spherical and deformed shape. The fact that $^{140}$Sm has a~very low lying $2_2^+$ state supports the notion of triaxiality. Indeed, earlier investigations of the level structure of $^{140}$Sm have found that triaxial rotor calculations with moment of inertia parameters obtained from Woods-Saxon calculations could reproduce the excitation spectrum reasonably well \cite{Kern1987}. As a~first approach, it seems therefore natural to interpret the electromagnetic matrix elements within the simple triaxial rotor model, although its applicability for a weakly deformed even-even nucleus is not evident. 

In the geometric model of Davydov and Filippov \cite{Davydov1958}, the excitation energy of the $2_2^+$ state is very sensitive to the degree of $\gamma$ deformation. The lowest energy is found for maximum triaxiality of $\gamma = 30^{\circ}$, where the $2_2^+$ state is at twice the energy of the $2_1^+$ state. Experimentally, the energy of the $2_2^+$ state in ${}^{140}$Sm is found to be even lower with $E(2_2^+)/E(2_1^+)=1.86$ \cite{Samorajczyk2015}, which suggests that $\gamma = 30^{\circ}$ should be used for the triaxial rotor model. For maximum triaxiality, the spectroscopic quadrupole moment $Q_s(2_1^+)$ should be zero, consistent with the present experimental result. Furthermore, the   relative $E2$ strength of the $2_2^+ \rightarrow 2_1^+$ transition is expected to be $B(E2;2_2^+ \rightarrow 2_1^+)/B(E2;2_1^+ \rightarrow 0_1^+)=10/7=1.43$, in good agreement with the experimental value of 1.52(25). This transition should only have a very small $M1$ component, which is again consistent with the present result and the $M1/E2$ mixing ratio extracted from the angular correlation measurement \cite{Samorajczyk2015}. 
The $2_2^+ \rightarrow 0_1^+$ transition is strictly forbidden in the triaxial rotor model for $\gamma = 30^{\circ}$. The experimental upper limit, $B(E2;2_2^+ \rightarrow 0_1^+) < 0.001$ $e^2$b$^2$, is indeed very low. The experimental energy ratio $E(4_1^+)/E(2_1^+) = 2.35$ is somewhat smaller than the value of 2.67 predicted by the Davydov-Filippov model, whereas the ratio
$B(E2;4_1^+ \rightarrow 2_1^+)/B(E2;2_1^+ \rightarrow 0_1^+) = 1.30(14)$ agrees within errors with the expected value of 1.43.

The good agreement between the experimental relative transition strengths and the predictions of the triaxial rotor model for $\gamma = 30^{\circ}$, in particular for the $2_2^+$ state, suggests that the triaxial degree of freedom is important to understand the structure of the low-lying states in $^{140}$Sm. It should be noted, however, that the relative $B(E2)$ strengths for surface vibrations in nuclei with $\gamma$-independent potential \cite{Wilets1956} are the same as for a~triaxial rigid rotor. Indeed, it was found earlier that models based on the Wilets-Jean Hamiltonian are able to describe transitional nuclei in this mass region reasonably well \cite{Rohozinski1974}.
Experimentally it is difficult to distinguish between $\gamma$-soft vibrational and $\gamma$-rigid rotational excitations. Coulomb excitation experiments are in principle able to distinguish between the two modes of excitation by providing complete sets of $E2$ matrix elements, which can be used to evaluate $E2$ invariants by summing over rotationally invariant zero-coupled products of $E2$ matrix elements \cite{Cline1986}. Higher-order products are not only sensitive to the centroid, but also to the fluctuation widths of the $E2$ invariants. In practice, the technique is very challenging as incomplete knowledge of the matrix elements can strongly affect the results in particular for the higher-order products. The $E2$ matrix elements obtained in this work are insufficient to attempt an analysis of rotational invariants.

The best signature to distinguish between $\gamma$ softness and $\gamma$ rigidity is the energy staggering of the states in the $K=2$ $\gamma$ band \cite{Zamfir1991}. In the extreme $\gamma$-unstable limit, the states of the $\gamma$ band form groups as $(2_{\gamma}^+)$, $(3_{\gamma}^+,4_{\gamma}^+)$, $(5_{\gamma}^+,6_{\gamma}^+)$ \cite{Wilets1956}, whereas for a~rotor with maximum triaxiality the staggering leads to a~grouping as 
$(2_{\gamma}^+,3_{\gamma}^+)$, $(4_{\gamma}^+,5_{\gamma}^+)$ etc.~\cite{Davydov1958}. The states of the $\gamma$ band beyond the $2_2^+$ state are unfortunately not known for $^{140}$Sm. The state at 1599~keV excitation energy that was tentatively assigned as the $3_{\gamma}^+$ state in an earlier work \cite{Kern1987} was shown to have spin-parity $I^{\pi} = 0^{(+)}$ in the recent angular correlation measurement \cite{Samorajczyk2015}. The identification of the $3_{\gamma}^+$ and $4_{\gamma}^+$ states would shed more light on the character of the axial asymmetry. Future Coulomb excitation experiments at higher beam energies and with heavier targets such as $^{208}$Pb will populate these states with higher cross sections than the present experiment.

\subsection{Algebraic models}

Although the matrix elements related to the $2_2^+$ state are well described by either $\gamma$-unstable or $\gamma$-rigid geometrical models, the discrepancies for the energies of the states indicate that these descriptions are too simple. The interacting boson approximation (IBA) was previously applied to describe transitional nuclei in this mass region \cite{Pascu2010}. The IBA parameters found to reproduce the experimental excitation energies of the lowest states place $^{140}$Sm in between the spherical vibrational and the triaxially soft rotational limits of the IBA, which correspond to the SO(6) and U(5) subalgebras, respectively. The transition between the SO(6) and U(5) dynamic symmetries has been interpreted as a~shape phase transition. At the critical point of this shape transition, the nuclear potential can roughly be approximated by a~five-dimensional infinite well, in which case the Bohr Hamiltonian can be solved analytically and the wave functions be expressed in terms of Bessel functions \cite{Iachello2000}. Both $^{134}$Ba and $^{128}$Xe have been identified as good examples for this so-called E(5) critical point symmetry \cite{Casten2001PRL85,Clark2004}. The predictions of the E(5) description for excitation energies and transition probabilities are parameter free except for overall scaling factors. To compare experimental data with the E(5) predictions, excitation energies and transition strengths are usually normalized to the energy of the $2_1^+$ state and the $B(E2;2_1^+ \rightarrow 0_1^+)$ value, respectively. 

The nucleus $^{140}$Sm has not been considered as a~candidate for E(5) critical point symmetry, mostly because of the previous assignment of the 990~keV level as a~$0^+$ state, which would be incompatible, and because experimental $B(E2)$ values were lacking. With the unambiguous identification of the states at 990 and 1599~keV as $2^+$ and $0^+$ states, respectively \cite{Samorajczyk2015}, and with the transition strengths obtained in this work, the hypothesis of $^{140}$Sm as a~candidate for E(5) critical point behavior can be evaluated. The excitation energies expected for a nucleus with E(5) symmetry are included in Fig.~\ref{fig:Schemes}.

The excitation energies of the $4_1^+$ and $2_2^+$ states are expected to be equal with energy ratios $E(4_1^+)/E(2_1^+)=E(2_2^+)/E(2_1^+) = 2.20$ \cite{Iachello2000}. The experimental ratios of $E(4_1^+)/E(2_1^+)=2.35$ and $E(2_2^+)/E(2_1^+)=1.86$ are somewhat larger and smaller than the E(5) predictions, respectively. For the $B(E2)$ values, the E(5) model finds ratios of $B(E2;4_1^+ \rightarrow 2_1^+)/B(E2;2_1^+ \rightarrow 0_1^+) = B(E2;2_2^+ \rightarrow 2_1^+)/B(E2;2_1^+ \rightarrow 0_1^+) = 1.56$ when using the $E2$ operator of the IBA \cite{Arias2001}. The experimental $B(E2)$ values obtained in the present work yield $B(E2;4_1^+ \rightarrow 2_1^+)/B(E2;2_1^+ \rightarrow 0_1^+) = 1.30(14)$ and $B(E2;2_2^+ \rightarrow 2_1^+)/B(E2;2_1^+ \rightarrow 0_1^+) = 1.52(25)$. It has been argued that the excitation energies and decay patterns of excited $0^+$ states provide a more stringent test of possible E(5) behavior than the properties of the $4_1^+$ and $2_2^+$ states \cite{Clark2004}. In the E(5) description, the states $I_{\xi\tau}^{\pi}$ are characterized by the quantum numbers $\xi$ and $\tau$, which are related to the zeros of the Bessel functions \cite{Iachello2000}. 
The $0^+$ state with $\xi=2$ and $\tau=0$, $0_{20}^+$, should be the lowest in energy with $E(0_{20}^+)/E(2_1^+)=3.03$, and it should decay predominantly to the $2_1^+$ state with a~branching ratio $B(E2;0_{20}^+ \rightarrow 2_1^+)/B(E2;0_{20}^+ \rightarrow 2_2^+) = 8.66$. The next $0^+$ state has quantum numbers $\xi = 1$ and $\tau = 3$ with $E(0_{13}^+)/E(2_1^+)=3.59$ and $B(E2;0_{13}^+ \rightarrow 2_2^+)/B(E2;0_{13}^+ \rightarrow 2_1^+) = 31.2$. Experimentally, the lowest excited $0^+$ state is found at an excitation energy of 1599~keV with $E(0_2^+)/E(2_1^+)=3.01$ and $B(E2;0_2^+ \rightarrow 2_2^+)/B(E2;0_2^+ \rightarrow 2_1^+) = 2.88$ \cite{Samorajczyk2015}. Two more states with tentative $0^+$ assignment were reported at excitation energies of 1629~keV ($E(0_3^+)/E(2_1^+)=3.07$) and 1933~keV ($E(0_4^+)/E(2_1^+)=3.64$). Both states were observed to decay only to the $2_1^+$ state without decay branch to the $2_2^+$ state \cite{Firestone1991}. Since only the $0_2^+$ state at 1599~keV has a~significant decay brach to the $2_2^+$ state, it could be associated with the $\xi = 1$, $\tau = 3$ state of the E(5) model, although the branching ratio would be expected to be even more in favor of the $2_2^+$ state. The $0_3^+$ state at 1629~keV would then be a~candidate for the $\xi = 2$, $\tau = 0$ state based on its excitation energy and decay to the $2_1^+$ state. 

Although $^{140}$Sm shows several features expected for a~nucleus with E(5) symmetry, there are also deviations from the expected energy and $B(E2)$ ratios. The assumption of a~nuclear potential that is independent of the deformation parameters $\beta$ and $\gamma$ is clearly too simple to describe the low-lying states in $^{140}$Sm. The excitation spectrum of $^{140}$Sm has been reasonably well described in the past using the IBA, including the presence of a~low-lying $2_2^+$ state below the $4_1^+$ state \cite{Kern1987}. To investigate whether the IBA can also reproduce the measured transition strengths, calculations were performed using the proton-neutron version of the model, IBA-2, with the same set of parameters as in previous investigations (parameter set 2 from Ref.~\cite{Kern1987}). The resulting excitation spectrum is indeed very similar to the one corresponding to E(5) symmetry ({\it c.f.} Fig.~\ref{fig:Schemes}), with the degeneracy of the multiplets only slightly lifted. However, this near degeneracy is not observed experimentally. 

Effective charges of $e_{\nu}=0.12$ and $e_{\pi}=0.13$ for neutrons and protons, respectively, were found in a~$\chi^2$ minimization to reproduce the experimental $B(E2;2_1^+ \rightarrow 0_1^+)$, $B(E2;4_1^+ \rightarrow 2_1^+)$, and $B(E2;2_2^+ \rightarrow 2_1^+)$ values. The values for the effective charges are similar to those found in previous investigations for this mass region \cite{Pascu2010}. The resulting $B(E2)$ values, which are included in Table~\ref{table:theory}, are in good agreement with the experimental values. Furthermore, also the IBA-2 calculations find a~vanishing $B(E2;2_2^+ \rightarrow 0_1^+)$ value, and the calculated branching ratio $B(E2;0_2^+ \rightarrow 2_2^+)/B(E2;0_2^+ \rightarrow 2_1^+) = 0.90$ is of the same order of magnitude as the experimental value of 2.88 for the $0_2^+$ state at 1599~keV. The calculated quadrupole moment for the $2_1^+$ state is $Q_s(2_1^+)=-0.106$ $e$b, consistent with the experimental value. It can be concluded that the interacting boson approximation is able to reproduce the excitation spectrum and transition strengths of $^{140}$Sm reasonably well. However, the model relies on parameter fitting and effective charges to achieve this agreement, and its predictive power is therefore limited. 

\subsection{Shell model}

Large scale shell model calculations have been carried out including the $0g_{7/2},1d_{5/2}, 1d_{3/2},2s_{1/2}$ and $1h_{11/2}$ valence space for both protons and neutrons above a closed $^{100}$Sn core. The effective GCN5082 \cite{Gniady} interaction is a realistic renormalized G-matrix with phenomenological monopole constraints, which has been used previously for the description of both low-lying and high-spin structures of nuclei with $50\le N,Z\le 82$ \cite{Sieja2009,Caurier2010,Petrache2015}.
The nucleus $^{140}$Sm with 12 valence protons and 28 valence neutrons
constitutes a numerical challenge for shell model calculations at the
limit of our present computing capability (m-scheme dimension
$\sim 1.86\times 10^{11}$), using the parallelized version of the Nathan
shell-model code~\cite{Caurier2005}.
To reduce the basis dimension we introduce a truncation scheme with respect to the $1h_{11/2}$ orbital excitations, allowing all neutron excitations and up to 4 proton excitations to the $1h_{11/2}$ orbital.

The energies of the low-lying states obtained in the shell model calculations are shown and compared to experimental data in Fig.~\ref{fig:Schemes}. 
In general, a rather good agreement is found for both yrast and yrare states.
The calculations reproduce the correct ordering of the states and in particular a low-lying $2_2^+$ state. Transition rates were calculated using effective charges of $0.65e$ and $1.65e$ for neutrons and protons, respectively. The results are included in Table \ref{table:theory} and the $B(E2)$ values show a good agreement with the experimental data. Several features found for the $2_2^+$ and $3_1^+$ states support their interpretation as members of a $K=2$ $\gamma$ vibrational band: The quadrupole moments for the $2_1^+$ and $2_2^+$ states are almost equal but of opposite sign, the quadrupole moment for the $3_1^+$ state is approximately zero, and it decays to the $2_2^+$ state via a strong $E2$ transition.

\begin{figure}[t!]
\centering
\includegraphics[width=0.9\columnwidth]{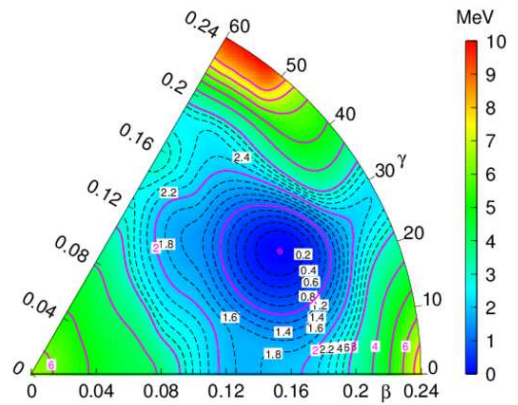}
		\caption{Potential Energy Surface for $^{140}$Sm from
                  constrained deformed Hartree-Fock minimisation in
                  the shell-model basis.}
	\label{figpes}
\end{figure}

Constrained Hartee-Fock calculations performed with the same shell-model Hamiltonian within the same valence space reveal the presence of a triaxial mininum at $\beta \sim 0.16$ and $\gamma=26.5^\circ$ as shown in Fig.~\ref{figpes}. The potential energy surface is consistent with both the geometrical interpretation of the data and the electromagnetic matrix elements obtained in the shell model calculations. The shell model calculations support the notion of triaxial deformation as the most salient feature of the $^{140}$Sm nucleus, with a well-developed $\gamma-$ soft band built on the $2_2^+$ state that is dominated by $2p2h$ proton excitations.

\subsection{Beyond mean field model}

\begin{figure}[ht!]
	\centering
        \includegraphics[width=0.8\columnwidth]{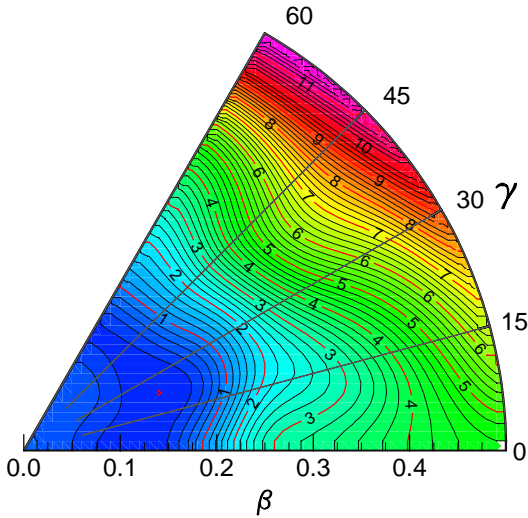}
	\caption{Potential energy surface for the ground state of $^{140}$Sm obtained in the CHFB calculations with the Gogny D1S interaction. 	\label{fig:ground_gogny}
}
\end{figure}

We have calculated the energies of the low-lying states and the transition strengths between them using microscopic calculations based on constrained Hartree-Fock-Bogoliubov (CHFB) theory using the Gogny D1S interaction~\cite{Decharge1980,Berger1991} and mapping to the five-dimensional collective Hamiltonian (5DCH) for quadrupole excitations at low energy. The method has been described in detail elsewhere \cite{Delaroche2010}, and results for the $B(E2;2_1^+ \rightarrow 0_1^+)$ values for the chain of neutron-deficient even-even samarium isotopes were presented in the context of the recent lifetime measurement in $^{140}$Sm \cite{Bello2015}, where it was demonstrated that the calculations are able to correctly describe the onset of quadrupole collectivity for the samarium isotopes below the $N=82$ shell closure. It should be noted that the CHFB+5DCH calculations contain no free parameters except for those specifying the phenomenological D1S interaction, which is globally used across the entire nuclear chart. Fig.~\ref{fig:ground_gogny} shows the potential energy surface obtained from the CHFB calculations.
A shallow minimum is found for quadrupole deformation parameters $\langle \beta \rangle = 0.17$ and $\langle \gamma \rangle =29^{\circ} $. The potential energy surface reveals considerable $\gamma$ softness. 

The excitation spectrum for $^{140}$Sm obtained in the CHFB+5DCH calculations for all positive-parity states up to the $6_1^+$ state is included in Fig.~\ref{fig:Schemes}. The general features of the excitation spectrum are well reproduced. The energies of the states in the ground-state band are overestimated by 10-15\%. It has been noted previously that the excitation energy of excited $0^+$ states are systematically overpredicted \cite{Delaroche2010}. This is also the case for $^{140}$Sm, where the discrepancy is approximately 30\%. The $2_2^+$ state is found to have predominant $K=2$ character, consistent with a $\gamma$-vibrational excitation. The calculations reproduce the relative position of the $2_2^+$ state slightly below the $4_1^+$ state. The $3_1^+$ and $4_2^+$ states can be interpreted as members of the $K=2$ $\gamma$ band. The energy spacing of the states in the $\gamma$ band is consistent with a $\gamma$-soft potential as evidenced by the potential energy surface. 

The calculated $B(E2)$ values are presented in Table \ref{table:theory}. The comparison with the experimental values shows very good agreement. The calculations find almost identical $B(E2)$ values for the $4_1^+ \rightarrow 2_1^+$ and $2_2^+ \rightarrow 2_1^+$ transitions and an almost vanishing $B(E2)$ value for the $2_2^+ \rightarrow 0_1^+$ transition, consistent both with the experimental results and the geometric interpretation of maximum triaxiality. 
The calculated quadrupole moment for the $2_1^+$ state is $Q_s(2_1^+)=-0.12$ $e$b, corresponding to a very small prolate deformation, in good agreement with the experimental value. The calculated quadrupole moment for the $2_2^+$ state, $Q_s(2_2^+)=+0.12$ $e$b, has the same magnitude and opposite sign compared to the quadrupole moment of the $2_1^+$ state, $Q_s(2_1^+)$, giving further support for the interpretation of the $2_2^+$ state as the head of a $\gamma$-vibrational band. As expected for such a band, the $3_1^+$ state decays predominantly to the $2_2^+$ state and has an almost vanishing quadrupole moment, $Q_s(3_1^+)=-0.01$ $e$b. As a whole, the calculations support the simple geometric picture of $^{140}$Sm as a nucleus with pronounced triaxiality and $\gamma$ softness and provide a fully microscopic foundation for this interpretation.

\section{Summary}\label{sec:summary}

A low-energy Coulomb excitation experiment to study electromagnetic transition probabilities and spectroscopic quadrupole moments in $^{140}$Sm was performed at the REX-ISOLDE facility at CERN. A quasi-pure beam of $^{140}$Sm was produced by proton-induced spallation of a primary tantalum target followed by resonant laser ionization. The radioactive ions were accelerated to an energy of $2.85 A$ MeV and scattered on a secondary $^{94}$Mo target. Scattered projectiles and recoiling target nuclei were detected in a highly segmented silicon detector at forward angles, while $\gamma$ rays were measured with the MINIBALL array of segmented HPGe detectors. The code GOSIA was used to extract transitional and diagonal electromagnetic matrix elements from the $\gamma$-ray yields observed as a function of scattering angle. By normalizing the yields to known reduced matrix elements for the $^{94}$Mo target and the known lifetime of the $2_1^+$ state, it was possible to determine several $B(E2)$ values for transitions between low-lying states and the spectroscopic quadrupole moment of the $2_1^+$ state in $^{140}$Sm. 

The experimental electromagnetic matrix elements were compared to the results of large-scale shell model calculations and to calculations based on constrained Hartree-Fock-Bogoliubov theory using the Gogny D1S interaction and mapping to a five-dimensional collective Hamiltonian. A clear picture emerges from the comparison between experimental and theoretical results that relate the observed structures in $^{140}$Sm to a weak quadrupole deformation with maximum triaxility of $\gamma \approx 30^{\circ}$ and significant $\gamma$ softness. The analysis of the excitation spectrum and the transition probabilities using the interacting boson model suggests that $^{140}$Sm exhibits many of the features expected for a nucleus with approximate E(5) critical point symmetry. 

To learn more about the degree of $\gamma$ softness it would be desireable to identify the states of the $\gamma$ vibrational band beyond the $2_2^+$ state and to measure transition probabilities connected to the states in the $\gamma$ vibrational band and to the excited $0^+$ states. With the higher beam energies provided by the new HIE-ISOLDE post-accelerator and the resulting higher cross sections, such measurements will become feasible in the future.

\section{Acknowledgements}\label{sec:acknowledgements}
We greatfully acknowledge the support of the ISOLDE collaboration and the technical teams at CERN.
This work was supported by the Research Council of Norway under project grants 213442, 210007 and 205528.
H.N.\ acknowledges support from the Helmholtz Association through
the Nuclear Physics Virtual Institute NAVI (N$^{\circ}$ VH-VI-417).
J.S.\ and P.N.\ acknowledge support from the National Science Center Poland through
decision DEC-2013/10/M/ST2/00427.
N.W.\ acknowledges the German BMBF under contracts 05P12PKFNE and 05P15PKCIA.
The work of P.V.D.\ was funded by FWO-Vlaanderen (Belgium), BOF KU Leuven (grant GOA/2010/010) and the Interuniversity Attraction Poles Programme initiated by the Belgian Science Policy Office (BriX network P7/12).
A.C.L.\ acknowledges support from ERC-STG-2014 under grant agreement no.\ 637686.

\bibliography{klintefjord}

\end{document}